\RequirePackage{fix-cm}
\documentclass[smallextended]{svjour3}       
\smartqed  
\usepackage{graphicx}
\begin{document}

\title{The effect of forcing on vacuum radiation}


\author{Katherine Brown        \and
    Ashton Lowenstein \and Harsh Mathur
}


\institute{K.Brown\at
               Physics Department\\
               Hamilton College \\
              198 College Hill Road \\
               Clinton, NY 13323\\              
               U.S.A. Tel.: +315-859-4585\\             
             \email{kjonessm@hamilton.edu}           
           \and
           A. Lowenstein \at
              Hamilton College, Clinton, NY 13323
 \and
 H.Mathur \at Case Western Reserve University, Cleveland, OH 44106
}

\date{Received: date / Accepted: date}

\maketitle

\begin{abstract}
Vacuum radiation has been the subject of theoretical study in both cosmology and
condensed matter physics for many decades. Recently there has been impressive progress
in experimental realizations as well. Here we study vacuum radiation when a field mode is
driven both parametrically and by a classical source. We find that in the Heisenberg picture
the field operators of the mode undergo a Bogolyubov transformation combined with
a displacement; in the Schr\"{o}dinger picture the oscillator evolves from the vacuum to a
squeezed coherent state. Whereas the Bogolyubov transformation is the same as would be obtained
if only the parametric drive were applied the displacement is determined by both the parametric
drive and the force. If the force is applied well after the parametric drive then the displacement
is the same as would be obtained by the action of the force alone
and it is essentially independent of $t_f$, the time lag between the application of the force and
the parametric drive. If the force is applied well before the parametric drive the displacement is 
found to oscillate as a function of $t_f$. This behavior can be understood in terms of quantum
interference. A rich variety of behavior is observed for intermediate values of $t_f$. The oscillations
can turn off smoothly or grow dramatically and decrease depending on strength of the
parametric drive and force and the durations for which they are applied.
The displacement depends only on the Fourier component of the force at a single resonant frequency
when the forcing and the parametric drive are well separated in time. However for a weak
parametric drive that is applied at the same time as the force we show that the displacement
responds to a broad range of frequencies of the force spectrum. Implications of our
findings for experiments are briefly discussed. 
\keywords{ Casimir effect \and Quantum field theory (low energy) \and Quantum gravity}
\end{abstract}

\section{Introduction}
\label{intro}
In a landmark paper, Casimir recognized that quantum vacuum fluctuations can produce measurable effects,
for example, an attractive force between two parallel conducting plates in a vacuum
\cite{casimir}. Subsequently, Moore \cite{moore} 
observed that motion of the conducting plates would
result in the generation of radiation from the vacuum, an effect dubbed the dynamical Casimir effect. 
Earlier, Parker \cite{parker} showed that in an expanding Universe, vacuum fluctuations 
would lead to particle production; according to the modern inflationary paradigm this mechanism
is the origin of structure in the Universe \cite{weinberg}. In his seminal work,
Hawking \cite{hawking} showed that even an apparently static system such as 
a non-rotating black hole will produce vacuum radiation due to its horizon. 
More recently, there have been significant advances on the experimental front:
the Casimir effect \cite{casimirexpt} and the dynamical Casimir effect \cite{superconduct}
have been unambiguously observed, and a laboratory analog of Hawking radiation proposed by Unruh \cite{unruh} 
has been claimed to have been observed in a Bose-Einstein
condensate \cite{bec}. 

A common element in much analysis of dynamical Casimir radiation is to assume that the system 
starts in the vacuum state. However in cosmology there is an ambiguity about what constitutes
the appropriate initial state and hence there has been an exploration of various alternative 
vacua as the initial state \cite{bunch}. Still earlier, in context of Hawking radiation from black holes, Wald
explored the consequences of starting the system in an excited state rather than the vacuum. 
He found that the quantum radiation was enhanced in a manner reminiscent of stimulated emission \cite{wald}. 
Here we wish to study the related but distinct effect of starting the system in the vacuum and examining
the quantum
radiation that results when the system is simultaneously excited 
parametrically (as in the dynamical Casimir effect) and also driven
directly by a classical source.
This analysis may be relevant to the experimental systems that are currently under investigation noted
above and discussed further in section \ref{sec:conclusion}. 
Another key motivation for this work is the observation of gravitational radiation from merging black holes \cite{ligo}. 
The gravitational radiation in this case is predominantly classical and reliable calculation of its magnitude
has only recently been achieved \cite{numerics}. 
In order to estimate the quantum contribution to radiation in a merger it would be necessary
to take into account the strong classical driving to which the gravitational radiation field is subject in these
events. Here for simplicity we primarily focus on a simple toy model of a single field mode. In 
section \ref{sec:fields} we sketch the generalization to coupled modes and a complete field theory. 

We assume that the oscillator experiences parametric driving (frequency $\omega$ varies in time approaching
the natural value $\omega_0$ asymptotically for $t \rightarrow \pm \infty$) and in addition an applied force $F(t)$. When treated individually the effect of each is well-known: 
in the Heisenberg picture, 
for a parametrically-driven oscillator the evolution of the ladder
operators is a Bogolyubov transformation, and for a forced oscillator the ladder operators
undergo simple displacement. Equivalently in the Schr\"{o}dinger picture the vacuum evolves into a squeezed vacuum state
under parametric driving and into a coherent state under forcing. Our principal finding is that
when the oscillator is driven parametrically as well as by a force then in the Heisenberg picture the
ladder operators undergo a Bogolyubov transformation combined with a  displacement (see
eq (\ref{eq:forcedsol})), or equivalently, in the Schr\"{o}dinger picture the vacuum evolves to a squeezed coherent
state. 

Furthermore, whereas the Bogolyubov transformation is found to be exactly the same as it would be in the
absence of forcing, the displacement is markedly influenced by the simultaneous
application of forcing and parametric driving. Only if the force is applied well after the 
parametric drive is the displacement the same as would have been obtained in the absence
of the parametric drive. If the force is applied well before the parametric drive, the final
state is the same as one would obtain if only the parametric drive were applied but the
starting state was an appropriately chosen coherent state rather than the vacuum. When the
parametric drive is applied at the same time as the force (the case of primary interest here)
it has a dramatic effect on the displacement. 

The behavior the displacement $\alpha$ as a function of the time lag between the 
force and the parametric drive $t_f$ is shown in fig \ref{fig:plotz}.
We show generically that for $t_f \rightarrow - \infty$ 
(force is applied much earlier than the parametric drive) $|\alpha|$ should undergo oscillations 
due to quantum interference; for $t_f \rightarrow \infty$ (force is applied much later than the
parametric drive) the displacement should approach a constant value controlled by the
Fourier component of the force at the frequency $\omega_0$. A rich variety of behavior
is seen at intermediate $t_f$. 
We find in the adiabatic limit (slowly varying $\omega(t)$) 
the oscillations turn off with a profile that may be described by a complementary 
error function. In the opposite abrupt limit too the
oscillations turn off smoothly and monotonically (fig \ref{fig:plotz} lower panel). However for 
other values of the parameters using an exactly soluble model we find that the oscillations
can grow dramatically even when they are negligible for $t_f \rightarrow - \infty$ (fig \ref{fig:plotz}
upper panel). The displacement depends only on the Fourier component of the force at frequency $\omega_0$
in the absence of parametric driving or when the parametric drive
and the force are well separated in time; however when the parametric drive is weak and applied at the
same time as the force we are able to show perturbatively that the displacement responds to a broad
range of frequencies in the driving force.

The remainder of this paper is organized as follows. In section \ref{sec:analysis} we 
present a general analysis of the model. We show that in the Heisenberg picture the field
operators will undergo a Bogolyubov transformation combined with a displacement and
derive general formulae for the transformation parameters and the displacement. We 
interpret these results in the Schr\"{o}dinger picture and extend the general analysis to multiple coupled
modes and field theory. In section \ref{sec:examples} we return to a single mode and analyze
a soluble model and various approximations that illustrate the rich variety of behavior that is
obtained due to the interplay of the forcing and the parametric drive. In section \ref{sec:conclusion} 
we discuss the implications of our results for experiments and open questions.

\section{Forced oscillator: general analysis}
\label{sec:analysis}

We consider a harmonic oscillator of mass $m$ whose natural frequency $\omega(t)$ varies in time; the time variation of the frequency represents the parametric driving that leads to vacuum radiation.
The frequency is assumed to approach the natural value $\omega_0$ asymptotically as $t \rightarrow \pm \infty$.
In addition the oscillator also experiences a time dependent force $F(t)$ that is also assumed to vanish 
asymptotically as $t \rightarrow \pm \infty$; the forcing corresponds to a classical source that drives the field. For the applications we envisage the oscillator represents
a single mode of a quantum field that is in some approximation decoupled from other degrees of freedom. 
Generally we will assume that the oscillator starts in the vacuum state $|0\rangle$ which is well defined 
as $t \rightarrow - \infty$ and we wish to determine the final behavior of the system as $t \rightarrow \infty$. Throughout the paper we adopt units where $\hbar =1$.

\subsection{Classical Analysis}
\label{sec:classical}
It is helpful to first solve the classical equation of motion
\begin{equation}
\frac{d^2 x}{d t^2} + \omega^2 (t) x= \frac{F(t)}{m}.
\label{eq:classical}
\end{equation}
For the case $F = 0$ eq (\ref{eq:classical}) has the form of a Schr\"{o}dinger equation for a particle 
scattering from a localized potential. Thus we can draw upon our intuition about the Schr\"{o}dinger
equation to deduce that there exists a solution with asymptotic behavior
\begin{eqnarray}
\xi (t) & \rightarrow & e^{-i \omega_0 t} \hspace{3mm} {\rm for} \hspace{3mm} t \rightarrow - \infty
\nonumber \\
& \rightarrow & A e^{-i \omega_0 t} + B e^{i \omega_0 t} \hspace{3mm} {\rm for} \hspace{3mm} t \rightarrow \infty
\label{eq:jost}
\end{eqnarray}
where $A$ and $B$ are scattering coefficients.
Since eq (\ref{eq:classical}) is real $\xi^\ast (t)$ 
constitutes a second independent solution. 
Note that these solutions have been chosen to satisfy Jost boundary conditions which are more
suitable for our purpose than the conventional scattering boundary conditions. 
It is evident from the Schr\"{o}dinger analogy that 
\begin{equation}
|A|^2 - |B|^2 = 1.
\label{eq:currentconservation}
\end{equation}
In section \ref{sec:examples} we solve eq (\ref{eq:classical}) approximately in various circumstances 
and exactly for a particular choice of $\omega^2 (t)$ and thereby obtain more explicit 
expressions for the coefficients $A$ and $B$ and for $\xi(t)$.

It is easy to verify that 
the solution that flows from the initial conditions $x (t_0) = x_0$ and $p(t_0) = p_0$ is
\begin{eqnarray}
\overline{x} (t) & = &  \frac{1}{2} \left( x_0 + i \frac{p_0}{m \omega_0} \right) e^{i \omega_0 t_0} \xi (t) 
\nonumber \\
& + & \frac{1}{2} \left( x_0 - i \frac{p_0}{m \omega_0} \right) e^{- i \omega_0 t_0} \xi^\ast (t).
\nonumber \\
\label{eq:xsolution}
\end{eqnarray}
Here we have assumed that $t_0$ is far in the past before the parametric drive was turned on.
Eq (\ref{eq:xsolution}) can be further simplified if we assume that $t$ is sufficiently far in the future 
that the parametric drive has been turned off. In that case we can use the $t \rightarrow \infty$ 
asymptotic form of $\xi(t)$ given in eq (\ref{eq:jost}). By differentiating eq (\ref{eq:xsolution}) 
one can then obtain $\overline{p} (t)$, the momentum for the solution that flows from the initial
conditions $(x_0, p_0)$. 

In order to analyze the effects of the force it is helpful to first consider the response $g(t, \tau)$  to an impulse
$F(t)/m = \delta( t - \tau )$ at time $\tau$. 
With the retarded boundary condition $g(t, \tau) = 0$ for $t < \tau$ the impulse response is given by
\begin{equation}
g(t, \tau) = \frac{ \theta( t - \tau) }{W} \left[ \xi (\tau) \xi^\ast (t) - \xi^\ast (\tau) \xi(t) \right]
\label{eq:green}
\end{equation}
where $\theta$ is the unit step function. The Wronskian $W = \xi (t) \xi^{\ast}$$' (t) - 
\xi'(t) \xi^\ast (t)$, where primes denote differentiation with respect to $t$. Since the Wronskian is a constant
independent of time we use  the $t \rightarrow - \infty$ behavior of $\xi$ to obtain 
$W = 2 i \omega_0$. 

Now by superposition the solution that flows from the initial condition $x(t_0) = x_0$ and
$p(t_0) = p_0$ is given by
\begin{equation}
x(t) = \overline{x} (t) + \frac{1}{m} \int_{t_0}^t d \tau \; g(t, \tau) F(\tau).
\label{eq:superpos}
\end{equation}
Here we assume that $t_0$ is far in the past before either the force or the parametric drive turn on. 
By differentiating eq (\ref{eq:superpos}) we then obtain the momentum $p(t)$ that flows from the 
initial conditions $(x_0, p_0)$ under the influence of both the force and the parametric drive. 

Eqs (\ref{eq:xsolution}) and (\ref{eq:superpos}) represent the solution to eq (\ref{eq:classical}) 
for a given initial condition. From the classical solution it is now easy to construct a complete 
solution to the corresponding quantum problem as discussed below. 

\subsection{Heisenberg picture}

\label{sec:heisenberg}

We turn now to the quantum problem. The quantum Hamiltonian corresponding to our model is
\begin{equation}
\hat{H} = \frac{1}{2 m} \hat{p}^2 + \frac{1}{2} m \omega^2(t) \hat{x}^2 - F(t) \hat{x}.
\label{eq:Hamiltonian}
\end{equation}
This leads to the Heisenberg equations of motion 
\begin{equation}
\frac{d}{d t} \hat{x} = \frac{1}{m} \hat{p}, \hspace{3mm}
\frac{d}{d t} \hat{p} = - m \omega^2 (t) \hat{x} + F(t).
\label{eq:heisenberg}
\end{equation}
First for simplicity let us assume $F = 0$. Because of their linearity
the solution to the quantum Heisenberg equations of motion can
be constructed using the classical solution eq (\ref{eq:xsolution}). 
We obtain
\begin{eqnarray}
\hat{x} (t) & = &  \frac{1}{2} \left( \hat{x}_0 + i \frac{\hat{p}_0}{m \omega_0} \right) e^{i \omega_0 t_0} \xi (t) 
\nonumber \\
& + & \frac{1}{2} \left( \hat{x}_0 - i \frac{\hat{p}_0}{m \omega_0} \right) e^{- i \omega_0 t_0} \xi^\ast (t)
\nonumber \\
\label{eq:xheisenberg}
\end{eqnarray}
and an analogous relation that expresses $\hat{p}(t)$ in terms of $\hat{x}_0$ and $\hat{p}_0$ 
from the classical expression for $\overline{p}(t)$. It is easy to verify that these expressions for
$\hat{x}(t)$ and $\hat{p}(t)$ satisfy the Heisenberg equations of motion and the appropriate 
initial conditions $\hat{x}(t_0) = \hat{x}_0$ and $\hat{p} (t_0) = \hat{p}_0$. 

For the quantum oscillator it is preferable to work with the creation operator $\hat{a} = \sqrt{m \omega/2} ( \hat{x} + i 
\hat{p}/m \omega)$ and the annihilation operator $\hat{a}^\dagger$. 
From the evolution of 
$\hat{x}$ and $\hat{p}$ we find
\begin{equation}
\hat{a} (t)  = u \hat{a}_0 + v \hat{a}^\dagger_0 \hspace{3mm} {\rm and} \hspace{3mm}
\hat{a}^\dagger (t) = u^\ast \hat{a}_0^{\dagger} + v^\ast \hat{a}_0. 
\label{eq:bogolyubov}
\end{equation}
Thus the evolution of the ladder operators is a Bogolyubov transformation with 
coefficients 
\begin{eqnarray}
u = A \exp[ - i \omega_0 (t - t_0) ] &\hspace{2mm} {\rm and} \hspace{2mm} &
v = B^* \exp [ - i \omega_0 (t + t_0) ] \nonumber \\
\label{eq:bcoefficients}
\end{eqnarray}
In eqs (\ref{eq:bogolyubov}) and (\ref{eq:bcoefficients}) 
we have assumed that $t_0$ is well before the parametric drive turns on and $t$ is well after. 

If we assume that the initial state of the system at $ t_0$ is the vacuum $|0\rangle$ defined by $a_0 | 0 \rangle = 0$
then in the Heisenberg picture of quantum mechanics the number of quanta excited at time $t$ is given by
\begin{equation}
\langle 0 | \hat{a}^\dagger (t) \hat{a} | 0 \rangle = | B |^2.
\label{eq:vacuum}
\end{equation}
This excitation of the system is the basic phenomenon of vacuum radiation.
Suppose following \cite{wald} we assume that the initial state of the system already has $n$ quanta then
\begin{equation}
\langle n | \hat{a}^\dagger (t) \hat{a} (t) | n \rangle - 
\langle n | \hat{a}^\dagger_0 \hat{a}_0 | n \rangle = (2 n + 1) | B |^2. 
\label{eq:lasing}
\end{equation}
This enhancement over the $n=0$ result
is a special case of the stimulated emission of vacuum radiation described by \cite{wald}. 
Another interesting initial state to consider is a coherent state $| \alpha \rangle$ defined by
$\hat{a}_0 | \alpha \rangle = \alpha | \alpha \rangle$ with $\alpha$ the complex coherent amplitude.
In this case we obtain
\begin{eqnarray}
\langle \alpha | \hat{a}^\dagger (t) \hat{a} (t) | \alpha \rangle -
\langle \alpha | \hat{a}^\dagger_0 \hat{a}_0 | \alpha \rangle & = &
(2 | \alpha |^2 + 1) | B |^2 
\nonumber \\
& + & 2 | A | | B | |\alpha|^2 \cos ( 2 \omega_0 t_0 + 2 \phi ).
\nonumber \\
\label{eq:coherent}
\end{eqnarray}
Here the phase $\phi$ is defined by the relation $\alpha^2 A B = |\alpha|^2 |A| |B| e^{i 2 \phi}$. Once again the first term on the right hand side corresponds to an enhancement compared to the excitation of
the vacuum $| 0 \rangle$ which is simply a coherent state with $\alpha = 0$. The second term shows a remarkable
oscillatory behavior in time due to quantum interference. Due to the Heisenberg dynamics the number of
quanta is uncertain at the final time even if it is definite initially; the interference is between states with
differing numbers of quanta present. We will see that this oscillatory behavior occurs
under many conditions when the oscillator is forced even if the initial state is the vacuum. 

Now let us include the force $F$ in our analysis. Once again the solution to the quantum Heisenberg equations
of motion can be constructed by adapting the classical solution eq (\ref{eq:superpos}) and its counterpart
for $p(t)$. Again it is preferable to work with the creation and annihilation operators $\hat{a}$ and $\hat{a}^\dagger$
rather than $\hat{x}$ and $\hat{p}$. The final result is
\begin{equation}
\hat{a} (t) = u \hat{a}_0 + v \hat{a}_0^\dagger + \alpha \hspace{3mm} {\rm and} \hspace{3mm}
\hat{a}^\dagger (t) = u^\ast \hat{a}_0^\dagger + v^\ast \hat{a}_0 + \alpha^\ast.
\label{eq:forcedsol}
\end{equation}
Thus the evolution of the ladder operators in this case is a Bogolyubov transformation together with
a constant  displacement $\alpha$. The coefficients $u$ and $v$ are still given by eq (\ref{eq:bcoefficients}) assuming
that $t_0$ is before the force or parametric drive are turned on and $t$ is after. The displacement 
$\alpha$ is given by 
\begin{equation}
\alpha =  \frac{ i e^{-i \omega_0 t} }{\sqrt{2 m \omega_0}}  
\int_{t_0}^t d \tau \; F(\tau)[A \;\xi^\ast (\tau) 
- B^\ast \xi (\tau) ].
\label{eq:displacement}
\end{equation}

It is useful to consider various special cases of this result. 
First suppose that there is no parametric driving. In that case $A = 1$ and
$B = 0$ and $\xi(\tau) = \exp ( - i \omega_0 \tau)$ for all time. Hence in this case
\begin{equation}
\alpha =  \frac{i e^{-i \omega_0 t}}{\sqrt{2 m \omega_0}}  
\tilde{f} (\omega_0) e^{i \omega_0 t_f} 
\label{eq:freeforced}
\end{equation}
and the number of quanta at time $t$ is given by $\langle 0 | \hat{a}^\dagger (t) \hat{a} | 0 \rangle = | \alpha |^2$.
Here we have taken the force to be $f(t - t_f)$ in the time domain. The force is assumed to be localized
about the time $t_f$. For explicit calculations below we will sometimes take the force to be a Gaussian
centered at $t_f$ with a sinusoidal modulation. 
Thus we arrive at the well-known result \cite{loudon},\cite{landaucm} 
that the excitation of the field oscillator is determined by $ \tilde{f} (\omega_0) $, the Fourier
amplitude of the force at the natural frequency of the mode, $\omega_0$.
In the Schr\"{o}dinger picture the state of the oscillator evolves from the vacuum $| 0 \rangle$ to the coherent
state $|\alpha\rangle$. 

Next suppose that the force is applied well before the parametric drive. In that case we can use
the $t \rightarrow - \infty$ behavior of $\xi (t)$ in order to evaluate $\alpha$. Making use of eq (\ref{eq:jost}) 
and \ref{eq:displacement}) we obtain
\begin{equation}
\alpha \approx  \frac{i e^{-i \omega_0 t}}{\sqrt{2 m\omega_0}} 
\left[ A \tilde{f}(\omega_0) e^{i \omega_0 t_f} 
- B^\ast \tilde{f}(\omega_0) e^{- i \omega_0 t_f} \right].
\label{eq:genearly}
\end{equation}
In this case too the displacement is determined entirely by 
the Fourier amplitude of the force at the natural frequency $\omega_0$.
Note that if we compute 
$| \alpha |^2$ it will have an interference term that oscillates with frequency $2 \omega_0$ as a function of
$t_f$. This oscillation is also manifest if we compute the number of quanta at late times 
\begin{eqnarray}
\langle 0 | \hat{a}^\dagger (t) \hat{a} (t) | 0 \rangle & = &  |B|^2 
+ \frac{ | \tilde{f}(\omega_0) |^2}{2 m \omega_0}
\left[ (2 |B|^2 +1) \right] 
\nonumber \\
& + & \frac{ | \tilde{f} ( \omega_0 ) |^2 }{ 2 m \omega_0 } 
\left[ 2 |A| |B| \cos ( 2 \omega_0 t_f + 2 \phi) \right]. \nonumber \\
\label{eq:earlyosc}
\end{eqnarray}
Here we have made use of eqs (\ref{eq:forcedsol}) and eq (\ref{eq:genearly})
and the phase $\phi$ is defined by $ A B [i f(\omega_0)]^2 = |A| |B| | f (\omega_0) |^2 e^{i 2 \phi}$.
The oscillation has a simple interpretation because of the temporal separation in the force and
the parametric drive. The force first causes the oscillator to go into a coherent state $| \alpha \rangle$ at time
$t_f$ where $\alpha$ is given by eq (\ref{eq:freeforced}) with $t \rightarrow t_f$; the subsequent parametric
drive then causes quantum interference between states with different numbers of quanta as in eq (\ref{eq:coherent}).
Indeed eq (\ref{eq:earlyosc}) is 
identical to eq (\ref{eq:coherent}) if we make the replacement $t_0 \rightarrow t_f$
and $\alpha \rightarrow i \tilde{f}(\omega_0)/\sqrt{2 m \omega_0}$. 

Another circumstance in which we can compute $\alpha$ is if the force is applied well after the parametric drive.
In this case we can use the late time asymptotic behavior of $\xi$. Making use of eqs (\ref{eq:jost}) and 
(\ref{eq:displacement}) we obtain
\begin{equation}
\alpha = \sqrt{ \frac{m}{2 \omega_0} } i e^{-i \omega_0 t} e^{i \omega_0 t_f} \tilde{f}( \omega_0 ).
\label{eq:genlate}
\end{equation}
Since in this case also the action of the force and the parametric drive occur separately it is not 
surprising that the displacement is the same as would be obtained if only the force were applied. 
In this case too the displacement is determined entirely by the Fourier amplitude of the force at the
natural oscillator frequency $\omega_0$. 

The analysis simplifies in the cases that the force and the parametric drive are temporally separated
or only one drive is present. However the main focus of this paper is on the new effects
that arise when the force and parametric drive are both simultaneously present. As we will see below
in this circumstance, among other new features, the displacement responds to a broad range of frequency
components of the force. In order to explore these features in the next section we analyze a number
of soluble models. It is worth noting that if we were interested only in parametric driving it would be
sufficient to determine the asymptotics of $\xi(t)$ or more precisely the coefficients $A$ and $B$.
However in order to calculate the displacement $\alpha$ due to the forcing 
it is necessary to determine the entire
trajectory $\xi(t)$ exactly or in some approximation.

\subsection{Schr\"{o}dinger Picture}
\label{sec:schrodinger}

Although in principle
everything can be worked out in the Heisenberg picture it is instructive
to examine the same dynamics in the Schr\"{o}dinger picture. 
In the Heisenberg picture the state remains fixed and operators evolve according to 
$a(t) = U^\dagger a_0 U$ where $a_0$ is the initial operator and $U$ is the evolution 
operator. In the Schr\"{o}dinger picture operators like $a$ remain fixed in time while the state
evolves according to $| \Psi(t) \rangle = U | \Psi(0) \rangle$. 

In order to analyze the dynamics of the states it proves useful to first evolve the
operators back in time according to the conjugate dynamics $a_c (t) = U a_0 U^\dagger$. It is not
difficult to verify that if $a(t)$ in the Heisenberg picture is given by eq (\ref{eq:forcedsol}) 
then the time-reversed dynamics is given by
\begin{equation}
a_c(t) = u^\ast a_0 + v^\ast a_0^\dagger + \alpha^\ast.
\label{eq:schrodingeroperators}
\end{equation}

Now it is easy to verify that if the initial state at time $t_0$ is the vacuum $|0 \rangle$, which is defined by
the condition $ a_0 | 0 \rangle$, then the state at time $t$, which is formally equal to $U | 0 \rangle$,
satisfies the condition
\begin{equation}
a_c (t) | \Psi (t) \rangle = 0.
\label{eq:squeeze}
\end{equation}
Eqs (\ref{eq:schrodingeroperators}) and eq (\ref{eq:squeeze}) fully determine the state $ | \Psi (t) \rangle$.
This state is a squeezed coherent state in the language of quantum optics \cite{loudonreview}. 
We see that for pure parametric driving ($\alpha = 0$) the final state obtained is a squeezed vacuum.
The effect of forcing is to produce a non-zero displacement $\alpha$ and lead to a final state that is
a squeezed coherent state. 

\subsection{Field theory formulation}

\label{sec:fields}

We now generalize the preceding results to a field theory with a large (possibly infinite) number of coupled modes. 
For notational simplicity we assume that there are $n$ coupled oscillators. Without loss of generality we may take
these modes to obey the equation of motion
\begin{equation}
\frac{d^2 x_i}{d t^2} + \omega_i^2 x_i + \sum_{j=1}^{n} \Omega_{ij}^2 (t) x_j = \frac{1}{m} F_i (t).
\label{eq:multimode}
\end{equation}
We assume that the modes are decoupled asymptotically and that the source also turns off asymptotically.
In other words we assume $\Omega_{ij}^2 (t) \rightarrow 0$ and $F_i(t) \rightarrow 0$ for $t \rightarrow \pm \infty$.

First let us analyze the case in which the field oscillators are only driven parametrically and $F = 0$. 
Evidently eq (\ref{eq:multimode}) then has $n$ independent solutions $\xi^{\mu}_i (t)$ with the label $\mu = 1, \ldots, n$. 
These solutions satisfy Jost boundary conditions
\begin{eqnarray}
\xi^\mu_i (t) & = & \sqrt{ \frac{\omega_0}{\omega_i} } \delta_{i \mu} e^{- i \omega_i t} \hspace{3mm} {\rm for} \hspace{3mm} 
t \rightarrow - \infty 
\nonumber \\
& = & \sqrt{ \frac{\omega_0}{\omega_i} } A_{i \mu}
e^{- i \omega_i t} + 
\sqrt{ \frac{\omega_0}{\omega_i} } B_{i \mu}
e^{+ i \omega_i t}
\hspace{3mm} {\rm for} \hspace{3mm} t \rightarrow \infty. 
\nonumber \\
\label{eq:fieldjost}
\end{eqnarray}
Here $\omega_0^2$ might represent the lowest of the frequencies $\omega_i$ (the ``mass gap'') or it might
be an arbitrarily chosen scale if the field theory we wish to analyze is gapless in the $n \rightarrow \infty$ limit. 
In addition there is a second set of $n$ independent solutions obtained by complex conjugation. 
The matrices $A$ and $B$ may be shown to satisfy
\begin{equation}
\sum_{i = 1}^n 
\left( A^\ast_{i \mu} A_{i \nu} - B^\ast_{i \mu} B_{i \nu} \right) = \delta_{\mu \nu}.
\label{eq:fieldunitarity}
\end{equation}
These solutions can be superposed to match any specified initial conditions exactly as in the the single mode case.

In order to incorporate the effect of the force we need the Green's function $G_{ij} (t, \tau)$ that obeys
\begin{equation}
\frac{ d^2 }{d t^2} G_{ij} (t, \tau) + \omega_i^2 G_{ij} (t, \tau) + \sum_{k=1}^n \Omega^2_{ik} G_{kj} (t, \tau) = 
\delta_{ij} \delta( t - \tau )
\label{eq:fieldgreen}
\end{equation} 
together with the boundary condition $G_{ij} (t, \tau) = 0$ for $t < \tau$. As in the single mode case the
desired Green's function can be constructed by use of the free solutions $\xi^\mu$. Thus
\begin{equation}
G_{ij} (t, \tau) = \frac{ \theta( t - \tau ) }{ W} \sum_{\mu = 1}^n 
\left[ \xi^\mu_j (\tau) \xi^{\mu \ast}_i (t) - 
\xi^{\mu \ast}_{j} (\tau) \xi^{\mu}_{i} (t) \right].
\label{eq:greenfield}
\end{equation}
Here the normalization factor $W = 2 i \omega_0$. 

Now making use of superposition we can write down a solution to eq (\ref{eq:multimode}) that flows 
from a specified initial condition by a straightforward generalization of the single mode analysis.
From this solution we can construct the transformation that connects the Heisenberg field operators at late times
to the initial field operators. The result is
\begin{equation}
a_i (t) = \sum_{\mu =1}^n \left[ U_{i \mu} a_\mu (t_0) + V_{i \mu} a_\mu^\dagger (t_0) \right] + 
\alpha_i
\label{eq:waldplus}
\end{equation}
Thus the evolution of the ladder operators is a Bogolyubov transformation together with a displacement
that is due to forcing. The Bogolyubov coefficients are given by 
\begin{equation} 
U_{i \mu}
= A_{i \mu} 
e^{- i \omega_i t} e^{i \omega_\mu t_0}
\hspace{3mm} {\rm and} \hspace{3mm}
V_{i \mu} 
= B_{i \mu}
e^{- i \omega_i t} e^{- i \omega_\mu t_0}
\label{eq:uvfield}
\end{equation}
while the displacement is
\begin{equation}
\alpha_i = \frac{ i e^{-i \omega_i t} }{ \sqrt{ 2 m \omega_0 } } 
\sum_{\mu, j} \sqrt{ \frac{ \omega_i }{\omega_\mu} } \int_{t_0}^t d \tau \; 
\left[ A_{i \mu} \xi^{\mu \ast}_j (\tau) 
- B^\ast_{i \mu} \xi^\mu_j (\tau) \right] F_j (\tau).
\end{equation}

Eq (\ref{eq:waldplus}) is the main result of this section. In the absence of forcing $\alpha_i = 0$ and eq (\ref{eq:waldplus})
reduces to the familiar result that parametric driving corresponds to a Bogolyubov transformation (see for example
eqs (2.7) and (2.8) of ref \cite{wald}). Eq (\ref{eq:waldplus}) generalizes these results to the case when both forcing
and parametric driving are applied.

\section{Forced oscillator: soluble models} 

\label{sec:examples} 

In order to gain further insight into the states produced by the combination of forcing and parametric
driving we now investigate four circumstances wherein the classical dynamics can be solved exactly
or in some approximation.

\subsection{The Sech potential}
\label{sec:exact} 

We choose the frequency to have the time dependence
\begin{equation}
\omega^2 (t) = \omega_0^2 + \frac{ \Omega^2 }{ \cosh^2 (t/T) }.
\label{eq:sechsquare}
\end{equation}
For this choice the classical equation of motion is exactly soluble in terms of hypergeometric functions. 
A closely related model was studied in \cite{bernard} corresponding to the choice 
$\omega^2(t) = \omega_0^2 + \Omega^2 \tanh^2 (t/T)$ in connection with vacuum radiation in a
two-dimensional expanding Universe. For our choice in eq (\ref{eq:sechsquare}) the solution to the
analogous Schr\"{o}dinger equation is presented in \cite{landauqm}. 
Transcribing that result we obtain 
\begin{eqnarray}
\xi (t) & = & \left( \frac{1 - \eta^2}{4} \right)^{- i \omega_0 T/2} 
\nonumber \\
& \times & _2F_1 \left[ - i \omega_0 T - s, - i \omega_0 T +s +1, 
- i \omega_0 T + 1; \frac{1}{2} (1 + \eta) \right].
\nonumber \\
\label{eq:hypergeometric}
\end{eqnarray}
Here $_2F_1$ is the hypergeometric function, $\eta = \tanh (t /T)$, and the parameter
\begin{equation}
s = \frac{1}{2} \left( \sqrt{1 + 4 \Omega^2 T^2} - 1 \right).
\label{eq:s}
\end{equation}
This solution has the asymptotic behavior given in eq (\ref{eq:jost}) with the coefficients
\begin{eqnarray}
A & = & \frac{ \Gamma( 1 - i \omega_0 T ) \Gamma( - i \omega_0 T )}{ \Gamma( - i \omega_0 T - s ) 
\Gamma( - i \omega_0 T + s + 1 ) },
\nonumber \\
B & = & \frac{i \sin \pi s}{ \sinh( \pi \omega_0 T) }.
\nonumber \\
\label{eq:absoluble}
\end{eqnarray}
It can be verified that $|A|^2 - |B|^2 = 1$ by use of the identity $\Gamma(z) \Gamma(1-z) = \pi/\sin(\pi z)$. 

Eq. (\ref{eq:vacuum}) indicates that if the system starts in the vacuum, in the absence of forcing the average
number of quanta generated by the vacuum process is $|B|^2$. From eq (\ref{eq:absoluble}) we see that 
in the adiabatic limit $\omega_0 T \gg 1$ the amount of vacuum radiation is exponentially suppressed. 
Note the oscillatory dependent of $|B|^2$ upon $s$. 
This is a well-known feature of the ${\rm sech}^2$ potential
which is known to be reflectionless for critical values of its amplitude. 

In order to calculate the effect of forcing we need to choose a particular form of the forcing function $F(t)$ 
and use eqs (\ref{eq:displacement}), (\ref{eq:hypergeometric}) and (\ref{eq:absoluble}). 
The integrals therein have to be evaluated numerically. 
We will return to a discussion of these results in connection with the
approximate solutions below. 

\subsection{The Born approximation}

If the frequency $\omega^2 (t)$ does not deviate significantly from the asymptotic value $\omega_0^2$
we can solve eq (\ref{eq:classical}) via perturbation theory. To this end it is useful to recast the problem
as an integral equation analogous to the Lipmann-Schwinger equation. 
We wish to solve eq (\ref{eq:classical}) with $F = 0$ subject to the Jost boundary conditions 
eq (\ref{eq:jost}). This is equivalent to the integral equation
\begin{equation}
\xi (t) = e^{- i \omega_0 t} - \int_{-\infty}^\infty d\tau\; g^{(0)} (t, \tau) [ \omega^2 (\tau) - \omega_0^2 ] \xi (\tau)
\label{eq:lippmann}
\end{equation}
where $g^{(0)}$ is the unperturbed Green's function that satisfies
\begin{equation}
\frac{d^2}{d t^2} g( t, \tau) + \omega_0^2 g (t, \tau) = \delta (t - \tau).
\label{eq:freegreen}
\end{equation}
A key difference from the conventional Lipmann-Schwinger equation is that we impose the boundary
condition $g^{(0)} (t, \tau) = 0 $ for $t < \tau$. Using the general result eq (\ref{eq:green}) we obtain
$g^{(0)} (t, \tau) = \theta( t - \tau) (1/\omega_0) \sin [ \omega_0 (t - \tau) ]$. 

Thus far we have rewritten the problem exactly. The first-order Born approximation for $\xi(t)$ is to
insert the zeroth-order approximation 
$\xi(\tau) \approx e^{-i \omega_0 \tau}$
on the right hand side of eq (\ref{eq:lippmann}). We obtain
\begin{equation}
\xi (t) = e^{-i \omega_0 t} - \int_{-\infty}^{t} d \tau \; \frac{1}{\omega_0} \sin [ \omega_0 (t - \tau) ] 
[ \omega^2 (\tau) - \omega_0^2 ] e^{- i \omega_0 \tau}.
\label{eq:bornapproximation}
\end{equation}
From eq (\ref{eq:bornapproximation}) we infer that
\begin{eqnarray}
A & = & 1 - \frac{i}{2 \omega_0} \int_{-\infty}^\infty d \tau \; [ \omega^2 (\tau) - \omega_0^2 ], 
\nonumber \\
B & = & \frac{i}{2 \omega_0} \int_{-\infty}^\infty d \tau \; [ \omega^2 (\tau) - \omega_0^2 ] e^{- i 2 \omega_0\tau }.
\nonumber \\
\label{eq:abborn}
\end{eqnarray}
Eqs (\ref{eq:bornapproximation}) and (\ref{eq:abborn}) constitute the first-order Born approximation for an
arbitrary $\omega^2 (t)$. 

In order to get more explicit expressions we consider $\omega^2(t)$ of the form in eq (\ref{eq:sechsquare}). 
Evaluating eq (\ref{eq:abborn}) for this choice yields
\begin{equation}
A = 1 - \frac{ i \Omega^2 T}{\omega_0} \hspace{3mm} {\rm and} \hspace{3mm}
B = \frac{ i \pi \Omega^2 T^2}{ \sinh ( \pi \omega_0 T ) }.
\label{eq:abbornsech}
\end{equation}
To leading order in $\Omega T$ these results match the exact result 
eq (\ref{eq:absoluble}). 
Finally making use of eqs (\ref{eq:displacement}) and eq (\ref{eq:bornapproximation}) we obtain 
for the displacement
\begin{equation}
\alpha = \frac{i e^{- i \omega_0 t} }{ \sqrt{2 m \omega_0} } 
\left[ A \tilde{F} (\omega_0) - B^\ast \tilde{F}^\ast (\omega_0) - \int_{-\infty}^\infty \frac{d \nu}{2 \pi}
\tilde{F}^\ast (\nu) R(\nu) \right].
\label{eq:broadband}
\end{equation}
Here $A$ and $B$ are given by eq (\ref{eq:abbornsech}) and the response function
\begin{equation}
R(\nu) = \frac{\Omega^2 T^2}{ \omega_0 - \nu } \frac{1}{ \sinh \left[ \pi (\omega_0 + \nu ) T / 2 \right] }.
\label{eq:rnu}
\end{equation} 

Eq (\ref{eq:broadband}) reveals a new effect that arises due to the interplay
of the forcing and the parametric drive. 
The first two contributions to $\alpha$ are determined entirely by the resonant Fourier component 
$\tilde{F} (\omega_0)$ of the force. However the third contribution in eq (\ref{eq:broadband})
depends on the full spectrum of the force weighted by the response function $R(\nu)$. 
$R(\nu)$ is peaked at $\nu \approx \pm \omega_0$ with a width of order $1/T$. Due to this 
term a force that is off-resonance can still produce a displacement when accompanied by a
parametric drive. 

\subsection{The Abrupt limit}

Consider the circumstance that the time scale $T$ over which the frequency
changes is very short compared to $\omega_0^{-1}$. In this case the change in the frequency can be
approximated by a delta function. For the particular case of eq (\ref{eq:sechsquare}) 
in the absence of forces we approximate eq (\ref{eq:classical}) by
\begin{equation}
\frac{d^2 x}{d t^2} + \omega_0^2 x + 2 \Omega^2 T \delta(t) x = 0.
\label{eq:abruptclassical}
\end{equation}
The coefficient of the delta function is chosen to match the weight $\int_{-\infty}^\infty d t [ \omega^2(t) - \omega_0^2 ]
= 2 \Omega^2 T$. 

Evidently in this case the solution with Jost boundary conditions is given by
\begin{eqnarray}
\xi(t) & = & e^{-i \omega_0 t} \hspace{3mm} {\rm for} \hspace{3mm} t < 0;
\nonumber \\
& = & e^{-i \omega_0 t} - \frac{2 \Omega^2 T}{\omega_0} \sin ( \omega_0 t )
\hspace{3mm} {\rm for} \hspace{3mm} t > 0.
\label{eq:abruptxi}
\end{eqnarray}
This corresponds to
\begin{equation}
A = 1 - i \frac{\Omega^2 T}{\omega_0} \hspace{3mm} {\rm and} \hspace{3mm} 
B = i \frac{\Omega^2 T}{\omega_0}. 
\label{eq:ababrupt}
\end{equation}
The solution (\ref{eq:abruptxi}) is obtained as usual by taking an appropriate superposition of plane waves on either
side of the delta function and imposing matching conditions across the origin. 
It is easy to verify that eq (\ref{eq:ababrupt}) 
is consistent with the exact solution of eq (\ref{eq:absoluble}) in the limit $T \rightarrow \infty$
whilst $\Omega^2 T$ is held constant.  

We assume that the force $F(t)$ is localized about a time $t_f$
and write it in the form $f(t - t_f)$. Using eqs (\ref{eq:displacement}), (\ref{eq:abruptxi}) and (\ref{eq:ababrupt})
we obtain 
\begin{equation}
\alpha = \frac{ i e^{-i \omega_0 t} }{\sqrt{2 m \omega_0}}
\left[ \tilde{f} (\omega_0) e^{i \omega_0 t_f} - \frac{2 \Omega^2 T}{\omega_0} \left\{
{\cal I} + {\rm Im} [ \tilde{f}^\ast (\omega)_0) e^{- i \omega_0 t} ] \right\} \right].
\label{eq:abruptalpha}
\end{equation}
Here 
\begin{equation}
{\cal I} = \int_0^\infty dt\; f(t - t_f) \sin (\omega_0 t).
\label{eq:cali}
\end{equation}
Together eqs (\ref{eq:ababrupt}) and (\ref{eq:abruptalpha}) allow us to answer all questions of interest
about the excitation of the oscillator by the combined parametric drive and forcing. 

\begin{figure}[h]
\includegraphics[width=3.25in]{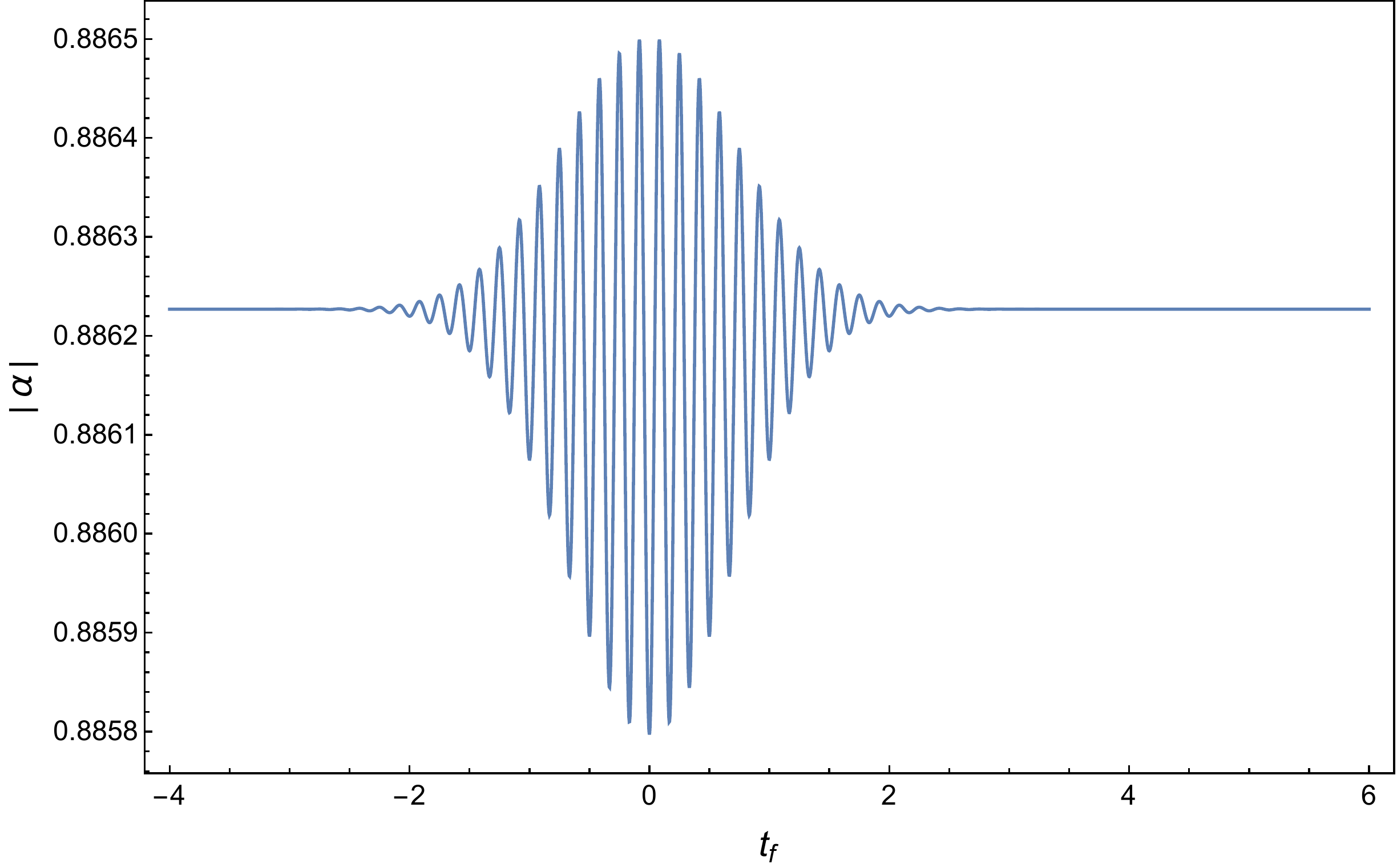}
\includegraphics[width=3.25in]{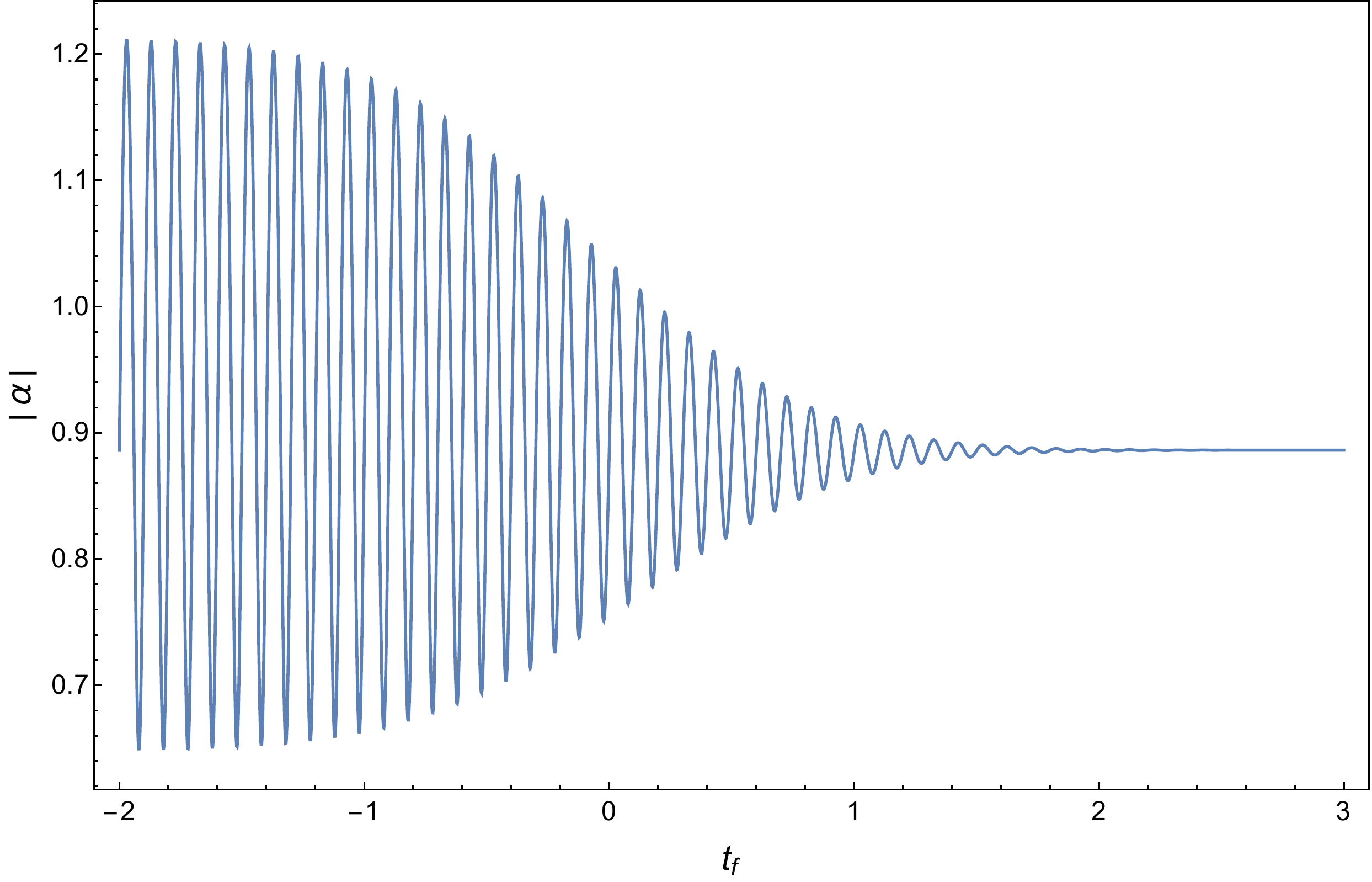}
\caption{Plot of the magnitude of the displacement $\alpha$ 
as a function of $t_f$, the time lag between the
application of the force and the parametric drive. 
The frequency $\omega(t)$ is assumed to be given by eq (\ref{eq:sechsquare}) 
and the force $F$ by eq (\ref{eq:force}). In the upper panel we choose $T_2 = 1$
and $T = 0.5, \Omega = 1, \omega_0 = 6 \pi, \omega_f = 6 \pi$. The lower
panel corresponds to the abrupt limit for which we choose $T_2 = 1$ and
$\Omega^2 T = 10, \omega_0 = 10 \pi, \omega_f = 10 \pi$. In both
cases on general grounds we expect to see oscillations for $t_f \rightarrow - \infty$
and a constant displacement for $t_f \rightarrow \infty$ [see eqs (\ref{eq:earlyosc})
and (\ref{eq:genlate})]. The asymptotic oscillations in the upper panel are suppressed
because $B \approx 0$ in this case. In the lower panel the oscillations are observed to
turn off smoothly as a function of $t_f$, 
behavior that is also found analytically in the opposite adiabatic limit.
In the upper panel by contrast the oscillations turn on only at intermediate times
illustrating the remarkable variety of possible behaviors.  
}
\label{fig:plotz} 
\end{figure}

Our principal result 
is that due to the combined effects of forcing and the parametric drive, the system goes into not just a squeezed vacuum but rather into a squeezed coherent state with 
displacement $\alpha$. Hence $|\alpha|^2$ is the key quantity of interest here. It is easy to verify
that in the limits $t_f \rightarrow \pm \infty$ when the forcing and the parametric drive are temporally
separate we recover the general results of section \ref{sec:heisenberg}. Now however we are in a 
position to examine what happens when the forcing and the parametric drive are applied simultaneously.
To this end we take the force to have the
form of a Gaussian of width $T_2$ that is modulated at a frequency $\omega_f$. Thus
\begin{equation}
F(t) = F_0 \cos [ \omega_f  (t - t_f) ] \exp \left[ - \frac{ (t - f_f)^2 }{T_2^2} \right].
\label{eq:force}
\end{equation}
For this form $\tilde{F} (\omega_0)$ and the integral ${\cal I}$ can be evaluated in closed form;
for the sake of brevity, these expressions are omitted. In fig \ref{fig:plotz} (lower panel) we plot
the resulting displacement as a function of $t_f$ for $T_2 = 1$ and $\Omega^2 T = 10, \omega_0  = 10 \pi,
\omega_f = 10 \pi$. For these parameters we see the expected asymptotic behavior, namely,
oscillations for $t_f \rightarrow - \infty$ and the approach to a constant displacement for 
$t_f \rightarrow \infty$. For intermediate $t_f$ we see that the oscillations turn off smoothly and
monotonically. This monotonic behavior is also found in the opposite adiabatic limit in 
section \ref{sec:adiabatic} below. However a variety of different behaviors are seen under 
other circumstances. For example in fig \ref{fig:plotz} (upper panel) we use the exact solution 
of section \ref{sec:exact} to plot the magnitude as a function of $t_f$ for $T_2 = 1$ and $T = 0.5, \Omega = 1, 
\omega_0 = 6 \pi$ and $\omega_f = 6 \pi$. For these values of the parameters the asymptotic
oscillations for $t_f \rightarrow - \infty$ have negligible amplitude since $B \approx 0$. However 
large oscillations are observed to turn on for intermediate values of $t_f$
when the force and parametric drive are 
applied simultaneously.

\subsection{The Adiabatic limit} 
\label{sec:adiabatic}
The adiabatic limit is of particular interest for experimental applications. Our objective is now to solve eq (\ref{eq:classical})
in the limit that $\omega^2(t)$ varies slowly. Formally we wish to take the limit $\omega_0 T \rightarrow \infty$. 
This is a subtle limit. If we interpret eq (\ref{eq:classical}) with $F = 0$ as a Schr\"{o}dinger equation, the problem
we wish to solve is that of reflection above a slowly varying barrier. The conventional WKB approximation
fails to capture the effect and yields $B \approx 0$. However by continuation in the complex plane one can
obtain the exponentially small value of the scattering coefficient \cite{landauqm}. For analysis of the forcing one needs
not only the scattering coefficients but rather the
entire solution $\xi(t)$. Considering its vintage this problem was solved relatively recently \cite{berry}
by a sophisticated use of divergent series.

Applying these techniques here we wish to solve
\begin{equation} 
\frac{d^2 x}{d t^2} + 
\left[ \omega_0^2 - \frac{ \Omega^2}{\cosh^2 (t/T)} \right] x = 0.
\label{eq:adiabatic}
\end{equation}
Note that we have changed the sign of the term with 
coefficient $\Omega^2$. The adiabatic analysis for the case with this sign is a bit
simpler and we focus on this case to avoid unneeded complications. 
We must assume that $\Omega^2 < \omega_0^2$ to ensure that the oscillator remains 
stable at all times. 

The exponentially small corrections to the conventional WKB approximation are 
controlled by the turning points at which $\omega^2 (t) = 0$. There are no turning points
for $t$ real but if we allow complex $t$ there are an infinity of turning points along the imaginary axis. 
Let $\varphi_1$ be the solution to $\cos \varphi = \Omega/\omega_0$ that lies in the range
$0 \leq \varphi_1 \leq \pi/2$. Then the two turning points closest to the real axis are
$t = \pm i \varphi_1 T$. There is an additional pair of turning points at $t = \pm i \varphi_2 T$
where $\varphi_2$ is the solution to $\cos \varphi = - \Omega/\omega_0$ that lies in the range
$\pi/2 \leq \varphi_2 \leq \pi$. Finally these four turning points are repeated periodically along the
imaginary axis at $t = \pm i \varphi_{1,2} T + 2 \pi i n T$ where $n$ is an integer. 

It is useful to rewrite eq (\ref{eq:adiabatic}) along the imaginary axis by making the substitution 
$t \rightarrow i s$. We obtain
\begin{equation}
\frac{d^2 x}{d s^2} + \left[ \frac{ \Omega^2}{\cos^2 (s/T) } - \omega_0^2 \right] x = 0.
\label{eq:imaginary}
\end{equation}
The conventional WKB solutions to this equation are real exponentials revealing that 
in the terminology of  \cite{berry} the segment of
the imaginary axis that connects the turning points $\pm i \varphi_1 T$ is a Stokes line
that intersects the real time axis. 

It now follows from ref \cite{berry} that the adiabatic solution to eq (\ref{eq:adiabatic}) with our preferred 
Jost boundary condition is 
\begin{eqnarray}
\xi (t) & \approx & \sqrt{ \frac{\omega_0}{\omega(t)} } \left \{ \exp \left[ - i \int_0^t d \tau \omega(\tau) \right] \right.
\nonumber \\
& + & \exp  \left. \left[ - \omega_0 T g \left( \frac{\Omega}{\omega_0} \right) \right] U(t) 
\exp \left[  i \int_0^t d \tau \omega (\tau) \right] \right \}.
\nonumber \\
\label{eq:berry}
\end{eqnarray}
where $g$ is defined below.
The upper line on the right hand side of eq (\ref{eq:berry}) corresponds to the conventional WKB solution.
The second line corresponds to the exponentially small backscattering. The coefficient $B$ is given by
\begin{equation}
B \approx \exp \left[ - \omega_0 T g \left( \frac{\Omega}{\omega_0} \right) \right] ,
\label{eq:reflection}
\end{equation}
while $U(t)$ is a smooth step function that goes from zero to one across the Stokes line with the universal form

\begin{equation}
U(t) = {\rm Erf} \left( \frac{t}{T_S} \right) \hspace{2mm} {\rm with} \hspace{2mm}
T_S = \frac{ \omega_0 T g ( \Omega/ \omega_0 ) }{ 2 \sqrt{ \omega_0^2 - \Omega^2} }. 
\label{eq:universal}
\end{equation}
The function $g$ is determined by the integral of $\omega(t)$ along the the Stokes line from the origin to the
nearest turning point 
\begin{equation}
\omega_0 T g \left( \frac{ \Omega }{ \omega_0 } \right) = \int_0^{\varphi_1 T} d s \; \sqrt{ \omega_0^2 - 
\frac{ \Omega^2}{ \cos^2 (s/T) } }
= \frac{\pi}{2} ( \omega_0 - \Omega ) T.
\label{eq:badiabatic}
\end{equation}


For illustration consider the displacement produced by a force $F(t) = f(t - t_f)$ 
that acts at time $t_f$ and has duration $T_2$. Assume for simplicity that $| t_f | \ll T$ so that the
force acts at the same time as the parametric drive and that the force is a brief impulse so that
$T_2 \ll T_S, T$. In this regime the integrals in eq (\ref{eq:displacement}) can be evaluated 
asymptotically and yield the result
\begin{eqnarray}
\alpha & = & 
\frac{ i e^{- i \omega_0 t} }{ \sqrt{2 m \omega_{{\rm eff}} } }  e^{ i \omega_{{\rm eff}} t_f } \tilde{f} (\omega_{{\rm eff}} )
\nonumber \\
& - & 
\frac{ i e^{- i \omega_0 t} }{ \sqrt{2 m \omega_{{\rm eff}} } } 
e^{- i \omega_{{\rm eff}} t_f} \tilde{f}^\ast (\omega_{{\rm eff}}) 
\exp \left[ - \omega_0 T g \left(
\frac{\Omega}{\omega_0} \right) \right] {\rm Erfc} \left(\frac{t}{T_S} \right). 
\nonumber \\
\label{eq:adiabaticalpha}
\end{eqnarray}
Here $\omega_{{\rm eff}} = \sqrt{ \omega_0^2 - \Omega^2}$ and ${\rm Erfc}$ is the
complementary error function. Eq (\ref{eq:adiabaticalpha}) reveals that if $|\alpha|$ 
is plotted as a function of $t_f$ there will be oscillations due to the interference
between the two terms. Note that the second term is modulated by a complementary
error function and vanishes as $t_f \rightarrow \infty$. The oscillations likewise
vanish as $t_f \rightarrow \infty$ with a complementary error function profile.

%

\section{Conclusion}

\label{sec:conclusion} 
In this paper we study the radiation produced when a field is excited both parametrically
and driven by a classical source. Normally in an experiment to measure the dynamical Casimir
effect all sources of radiation besides the parametrically excited vacuum radiation are minimized
and the field is not driven classically. However our finding that the parametric drive has a strong
effect on the radiation field when both kinds of excitation are applied suggests that it may be
possible to extract signatures of the 
quantum effects of the parametric drive even in experiments where classical sources are present.

Cavitation of bubbles in superfluid helium provides a possible realization of this physics.
The motion of the bubble wall would parametrically excite phonons in the fluid much as a moving
mirror excites photons \cite{moore}. Moreover, unlike a moving mirror, a moving bubble wall is
a strong classical source of acoustic radiation \cite{landaufluid}. Cavitating bubbles might even be able to achieve
supersonic flow \cite{putterman} leading to the formation of a sound horizon \cite{unruh}. 
Although bubbles in superfluid helium are well-studied \cite{heliumreview} these are likely challenging
experiments and it would be desirable in future work to develop an optimum experimental design. 

In ref \cite{superconduct} the dynamical Casimir effect was realized by terminating
a transmission line with a SQUID. Changing the flux in the SQUID modifies the boundary
condition at the termination effectively mimicking a moving mirror. The flux is varied periodically
coupling electromagnetic modes in pairs and leading to the formation of two-mode squeezed
states \cite{loudonreview}. In this case it is feasible to drive 
the transmission line with
a weak classical voltage source. Designing an unambiguous signature of the quantum
effects of the parametric drive on the resulting radiation field is therefore an interesting question 
for future work.

A primary motivation for this work is to develop an estimate of the quantum contribution
to gravitational radiation produced by astrophysical phenomena such as black hole mergers
that are accessible to LIGO and its planned successor gravitational wave observatories. 
Mergers are strong classical sources of gravitational radiation and in this case quantum
effects, if at all detectable, necessarily have to be observed against this background.

%



\end{document}